\documentclass[amsmath,notitlepage,twocolumn,prl,superscriptaddress,fixfloats]{revtex4-1}
\RequirePackage{varioref}
\RequirePackage{amssymb}
\RequirePackage{bm}
\RequirePackage{graphicx}
\RequirePackage{color}

\RequirePackage[T1]{fontenc}
\RequirePackage[utf8]{inputenc}
\RequirePackage{lmodern}

\newcommand\UNSW{Centre for Quantum Computation and Communication Technology, School of Electrical Engineering and Telecommunications, UNSW Sydney, Sydney NSW 2052, Australia}
\newcommand\UNSWphysics{School of Physics, UNSW Sydney, Sydney NSW 2052, Australia}
\newcommand\HRL{HRL Laboratories, LLC,
                3011 Malibu Canyon Rd., Malibu, CA 90265, USA}
\newcommand\UM{Centre for Quantum Computation and Communication Technology, School of Physics, University of Melbourne, Melbourne VIC 3010, Australia}
\newcommand\Keio{School of Fundamental Science and Technology, Keio University, 3-14-1 Hiyoshi, 223-8522, Japan}

\newcommand\storyboard[1]{\blue{#1}}

\newcommand{\mathcommand}[3][0]{\newcommand{#2}[#1]{\ensuremath{#3}}}
\mathcommand\sinc{\text{sinc}}
\renewcommand{\vec}[1]{\ensuremath{\text{\textbf{#1}}}} % Simply a bold charactor.
\newcommand{\be}{\begin{equation}}
\newcommand{\ee}{\end{equation}}
\newcommand{\blue}{\textcolor[rgb]{0,0,0.7}}

\newcommand{\ts}[2]{{#1}_{\textnormal{#2}}} %A math symbol with a text subscript

\mathcommand[1]{\smallket}{|#1\rangle}
\mathcommand[1]{\smallbra}{\langle #1|}
\mathcommand[1]{\bigket}{\bigl|#1\bigr\rangle}
\mathcommand[1]{\bigbra}{\bigl\langle #1\bigr|}
\mathcommand[1]{\biggket}{\biggl|#1\biggr\rangle}
\mathcommand[1]{\biggbra}{\biggl\langle #1\biggr|}
\mathcommand[1]{\ket}{\left| #1 \right\rangle}  %a Ket
\mathcommand[1]{\lket}{\bigl| #1 \bigr)}  %a Liouville Ket
\mathcommand[1]{\bra}{\left\langle #1 \right|}  %a Bra
\mathcommand[1]{\lbra}{\bigl( #1 \bigr|}  %a Bra
\mathcommand[1]{\bbra}{\biggl\langle #1 \biggr|}
\mathcommand[1]{\bket}{\biggl| #1 \biggr\rangle}
\newcommand{\refeq}[1]{Eq.~\eqref{#1}}

\renewcommand\storyboard[1]{} % comment out to restore storyboard (blue) notes
\begin{document}

\newcommand\thistitle{Controllable freezing of the nuclear spin bath in a single-atom spin qubit}
\author{Mateusz T. M\k{a}dzik}\affiliation\UNSW
\author{Thaddeus D. Ladd}\affiliation\UNSWphysics\affiliation\HRL
\author{Fay E. Hudson}\affiliation\UNSW
\author{Kohei M. Itoh}\affiliation\Keio
\author{Alexander M. Jakob}\affiliation\UM
\author{Brett C. Johnson}\affiliation\UM
\author{David~N.~Jamieson}\affiliation\UM
\author{Jeffrey C. McCallum}\affiliation\UM
\author{Andrew S. Dzurak}\affiliation\UNSW
\author{Arne Laucht}\affiliation\UNSW
\author{Andrea Morello}\affiliation\UNSW
\date{\today}

\title\thistitle
\begin{abstract}
The quantum coherence and gate fidelity of electron spin qubits in semiconductors is often limited by noise arising from coupling to a bath of nuclear spins. Isotopic enrichment of spin-zero nuclei such as $^{28}$Si has led to spectacular improvements of the dephasing time $T_2^*$ which, surprisingly, can extend two orders of magnitude beyond theoretical expectations. Using a single-atom $^{31}$P qubit in enriched $^{28}$Si, we show that the abnormally long $T_2^*$ is due to the controllable freezing of the dynamics of the residual $^{29}$Si nuclei close to the donor. Our conclusions are supported by a nearly parameter-free modeling of the $^{29}$Si nuclear spin dynamics, which reveals the degree of back-action provided by the electron spin as it interacts with the nuclear bath. This study clarifies the limits of ergodic assumptions in analyzing many-body spin-problems under conditions of strong, frequent measurement, and provides novel strategies for maximizing coherence and gate fidelity of spin qubits in semiconductors.

\end{abstract}
\maketitle

\section{Introduction}

\storyboard{Silicon qubits are cool but have nuclear noise}
Electron spin qubits in semiconductors are prominent candidates for building blocks of scalable quantum computers, thanks to a combination of small physical size, long quantum coherence times, and potential manufacturability using industry-standard processes \cite{Hanson2007a,Zwanenburg2013}. Demonstrating a universal set of quantum gates with fidelities beyond the fault-tolerance threshold remains the main focus of current research in the field.

Microscopically, one of the main sources of decoherence and gate errors is the coupling between the electron spin qubit and the mesoscopic bath of nuclear spins present in the host semiconductor material. The fluctuating polarization of the nuclear spin bath produces an effective magnetic field noise, which results in a random component of the instantaneous value of the qubit resonance frequency. This randomness affects the qubit coherence time, since the precise rate of phase accumulation is no longer accurately known, and the quantum gate fidelities, since the frequency of the classical control fields may be off-resonance with the instantaneous precession frequency of the spin qubit.

Silicon --- the material underpinning all of the modern microelectronic industry --- also has the key property of possessing a large natural abundance of the spin-zero nuclear isotope $^{28}$Si, with only 4.7\% $^{29}$Si nuclei carrying a spin $I=1/2$.
Isotopic enrichment methods compatible with wafer-scale fabrication have provided $^{29}$Si concentrations of order 500 -- 800~ppm \cite{Itoh2014,Sabbagh2018}. As was known for decades \cite{Gordon1958}, and more recently studied in detail in ensemble experiments on $^{31}$P donors in silicon \cite{Tyryshkin2003,Tyryshkin2012}, the electron spin coherence (as measured by a Hahn echo) in such enriched $^{28}$Si substrates is significantly improved from the values found in natural silicon. There is excellent agreement between measurements of electron spin Hahn echo decay in donor ensembles, and their predicted behavior based on the theoretical description of $^{29}$Si nuclear spin dynamics \cite{Witzel2007,Witzel2010}.

A series of experiments has been carried out in recent years on single-donor ion-implanted spin qubit devices in enriched silicon with 800 ppm residual $^{29}$Si. These experiments have qualitatively confirmed that the enrichment of spin-zero isotopes improves spin coherence times \cite{Muhonen2014} and quantum gate fidelities \cite{Muhonen2015,Dehollain2016b}. The Hahn echo decay time, $T_2^{\rm Hahn} \approx 200$~$\mu$s in natural silicon \cite{Pla2012}, was shown to increase by a factor of $\sim5$ in enriched $^{28}$Si \cite{Muhonen2014}. However, the ``pure dephasing'' time $T_2^*$ of single-donor electron spins went from $T_2^* \approx 55$~ns in natural silicon, to $T_2^* \approx 270$~$\mu$s in 800 ppm material --- an improvement of over three orders of magnitude. Strikingly, this surpasses by nearly two orders of magnitude what would be predicted assuming an ergodic behavior of the spin polarization of the residual $^{29}$Si nuclei.

In electron spin qubit systems in GaAs and Si/SiGe, a reasonably good prediction for the $T_2^*$ value is obtained by the following simple consideration \cite{vanvleck1948,Coish2004}.  Suppose that, over a sequence of experiments on an electron spin qubit, the system samples with equal probability every possible configuration of the nuclear spins coupled to the qubit.  The total electron-nuclear interaction is described by the Hamiltonian
\begin{equation}
\label{HHF}
\ts{H}{HF}=\sum_j A_j I_j^z S^z,
\end{equation}
where $A_j$ is the hyperfine interaction (dominated by the Fermi contact term) between the electron and the $j$-th nucleus, $I_j^z$ is the $z$-component of the spin operator for nucleus $j$, and $S^z$ is the $z$-component of the spin operator for the electron. In a semiclassical picture, the hyperfine Hamiltonian above can be thought of resulting in a longitudinal magnetic field $B_{\rm HF}$ coupling to the electron (Overhauser field).  This Hamiltonian assumes a large applied magnetic field $B_0$ in the $z$-direction, which allows neglecting the terms which do not conserve Zeeman energy.  If we presume a sufficiently long averaging time to ensure that the fluctuations in nuclear bath eventually lead the system to sample all possible nuclear spin configurations, then the variance of the Overhauser field $B_{\rm HF}$ is found as the sum of the variances of equal binomial random variables for each nuclear spin, giving a total variance $\sigma_\omega$ in the Larmor frequency $\omega_e$ of the electron spin:
\begin{equation}
\label{ergodic_sigma}
\sigma_\omega^2=\frac{2}{(T_{2\infty}^*)^2}=\frac{I(I+1)}{3}\sum_j A_j^2.
\end{equation}

As a result, electron coherences will decay as $\exp[-(\sigma_\omega t)^2/2]$.   The ergodic limit for $T_2^*$, here notated $T_{2\infty}^*$, is therefore $\sqrt{2}/\sigma_\omega$ according to \refeq{ergodic_sigma}. The effect of the fluctuating nuclear bath can be captured by describing it via spin-diffusion from the dipole-dipole interaction, leading to a local power spectral density proportional to $1/\omega^2$~\cite{Abragam1961,Reilly2008,Witzel2014}.

In the case of a $^{31}$P donor in silicon, the hyperfine constants $A_j$ are well-known and tabulated in Ref. \cite{Ivey1975a}.
Using these tabulated values and evaluating a large ensemble of random $^{29}$Si placements at 800 ppm concentration, we find that $\sigma_\omega^2$ is log-normal distributed as $10^{0.4 \pm 0.7}$~(rad/sec)$^2$.  The probability of a random placement of nuclear spins providing $T_{2\infty}^*>200$~$\mu$s is therefore less than 1 in 50 million. This begs the question as to why observed dephasing times are two orders of magnitude larger than a reasonable value for $T_{2\infty}^*$. The question is particularly pressing in view of the fact that, also in the case of pure dephasing times (or its inverse, the inhomogenous linewidth) theory and experiments agree very well in the case of spin ensembles \cite{Abe2010}. It is also pressing because the fidelity of 2-qubit logic operations, or their control resources for noise compensation, depend more crucially on the pure dephasing time than on the Hahn echo time~\cite{huang2019}.

Understanding the microscopic reason why the single-spin experimental results consistently deviate from the ergodic model is the aim of this work. We present a simple experiment, designed to provide a clear insight into what freezes the dynamics of the $^{29}$Si spin bath. The experiment also shows the ability to control the dynamics of the nuclear bath, and in turn approaches the spin dephasing values expected from an ergodic assumption. We then provide a detailed and nearly parameter-free theoretical model that puts our observation in a quantitative framework, and reveals important details on the interplay between qubit measurement and nuclear bath dynamics. By using the $^{31}$P donor system in 800 ppm enriched silicon, the small (of order 10) number of $^{29}$Si nuclei involved allows us to treat the problem in a brute-force, numerically accurate way.

Importantly, our work provides insights into decoherence processes that could not, even in principle, be obtained from experiments on spin ensembles. By definition, the local Overhauser field averaged over a large spin ensemble will exhibit statistics that reflect the distribution of all possible $^{29}$Si nuclear spin configurations (unless some hyperpolarization method is applied). Therefore, only a single-spin experiment can unveil the precise statistics and timescales over which the Overhauser field actually explores the whole range allowed by the nuclear spin concentration. Additionally, our theoretical model includes explicitly the back-action on the nuclear bath of the electron spin used by measurement. This turned out to be a key ingredient to fully describe the qubit and bath dynamics, also unavailable from ensemble magnetic resonance experiments.

\begin{figure*}
\includegraphics[width=\textwidth]{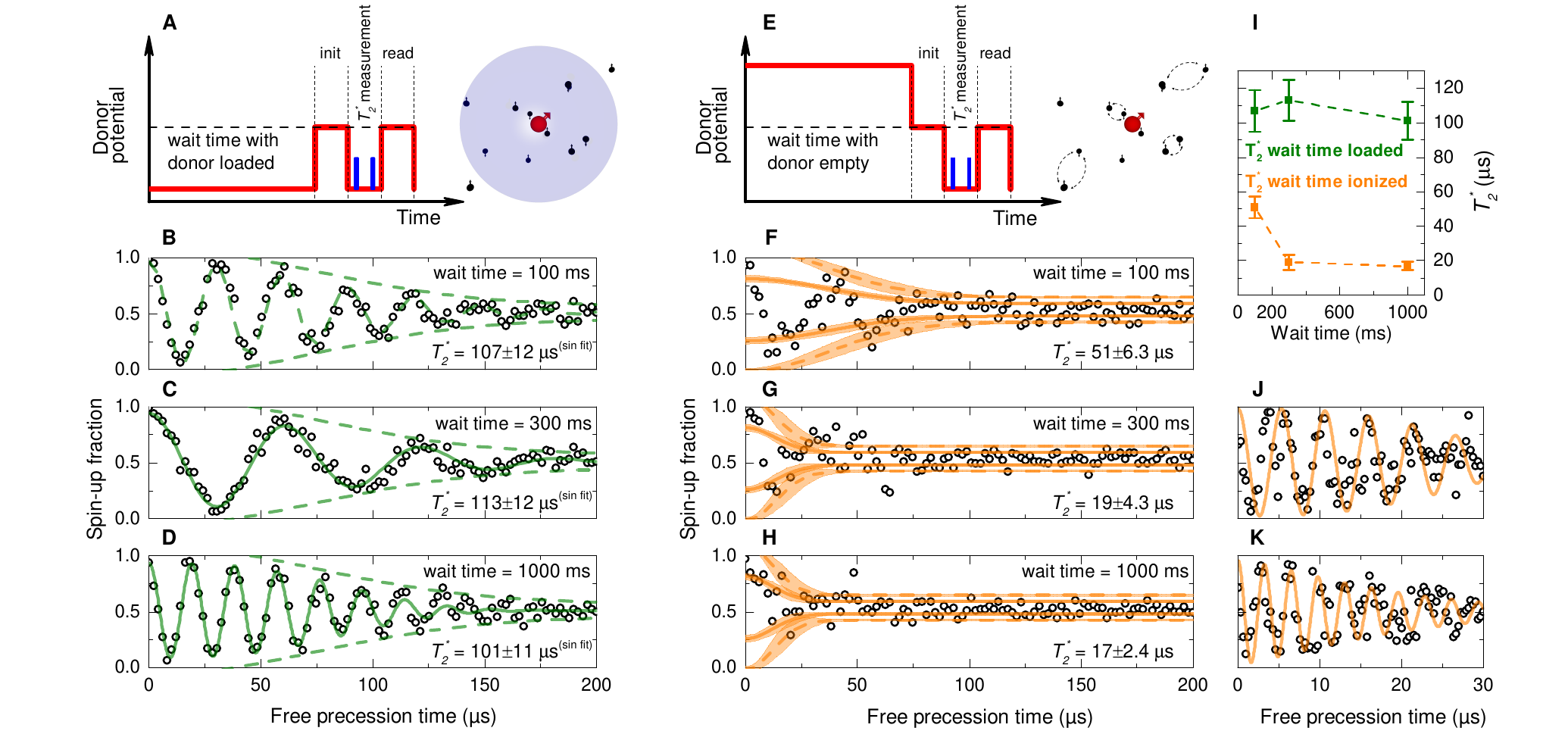}
\caption{\textbf{Experiment to quantify the impact of the donor electron on nuclear dynamics and example $T_2^*$ measurements}.  Panels \textbf{a}-\textbf{d} are for a loaded, or neutral donor, for most of the duration of the experiment; panels \textbf{e}-\textbf{h} are for an unloaded, or ionized donor. \textbf{a} and \textbf{e} show the pulse sequence applied; the donor potential is biased for some time $T_{\text{wait}}$ in its corresponding state; an initialization time of $T_{\text{load}}$=100~ms is then introduced to load an electron, in the empty case, followed by a Ramsey sequence in which coherent electron precession is framed by electron-spin-resonant (ESR) $\pi/2$ pulses.  The electron-spin is then read via spin-dependent tunneling during the ``read." interval.  \textbf{b-d} and \textbf{f-h} show the results of averaged single-shot spin measurements for three different values of $T_{\text{wait}}$ in each case.  In the loaded case, \textbf{b-d}, $T_2^*$ is longer and data is fit as a Gaussian-decaying sinusoid.  In the unloaded case, \textbf{f-h}, a Gaussian-decaying standard deviation $\sigma(\tau)$ at each $\tau$ is estimated to capture the envelope decay, and the single-$\sigma(\tau)$ (solid) and $2\sigma(\tau)$ (dashed) curves are shown, shaded via the standard-error confidence interval (see Materials and Methods section).  A decaying sinusoid may also be fit to obtain the same result using a larger ESR frequency offset and finer $\tau$ resolution, as shown in \textbf{j}-\textbf{k}.  Panel \textbf{i} indicates the clear difference in $T_2^*$ vs. $T_{\text{wait}}$ between the unloaded and ionized cases; error bars indicate standard error from nonlinear curve fitting.}
\label{schematic}
\end{figure*}

\section{Results}

\subsection{Experimental evidence of nuclear freezing by hyperfine interactions}

The design of our experiment aims at testing the ``nuclear freezing'' hypothesis, which postulates the existence of a ``frozen core'' of nuclear spins close to the electron spin \cite{Khutsishvili1962,Guichard2015}. The dynamics of such nuclear spins is frozen by the presence of electron-nuclear hyperfine interactions which vary strongly at each lattice site, causing the nuclei to have significantly different energy splittings. As a consequence, energy-conserving nuclear flip-flop processes are strongly suppressed, since the nuclear dipole-dipole interaction is much weaker than the energy difference caused by the local variation of hyperfine couplings. Some aspects of the physics of nuclear freezing have been studied with ensemble NMR experiments in the context of nuclear spin diffusion~\cite{Wald1992}, and signatures of these effects appear in combined optical and magnetic resonance studies of single defects~\cite{Bradley2019}, but here we aim at providing direct evidence of the impact of nuclear freezing on the dephasing of an electron spin qubit using controlled electron occupation.

The key novelty of our experiment is the deliberate use of our ability to ionize single phosphorus donors in gated nanostructures \cite{Pla2013}, in order to ``unfreeze" nuclear spins on demand.  Figures~\ref{schematic}a and \ref{schematic}e show the basic schematic of the experimental protocol, in which we perform measurements of $T_2^*$ using the Ramsey technique while maintaining the donor in different ionization states. First, the donor system is initialized in the ground electron $\ket{\downarrow}$ state. The $^{31}$P nuclear spin is set in the $\ket{\Uparrow}$ state and remains unchanged during the entire duration of the experiment. From here onwards the $^{31}$P nucleus will be ignored, since it plays no role other than providing a constant hyperfine frequency shift, i.e. causing the nominal electron spin Larmor frequency to be $\omega_{\rm e} = \gamma_e B_0 + A_{\rm P}/2$, where $\gamma_{\rm e}/2\pi \approx 28$~GHz/T is the electron gyromagnetic ratio, and $A_{\rm P}/2\pi \approx 97$~MHz is the hyperfine coupling to the $^{31}$P nucleus~\cite{Laucht2015,Muhonen2017}. Then, a microwave electron-spin-resonance (ESR) $\pi/2$ pulse at frequency $\omega_{\rm MW}$ establishes an electron-spin coherence of the form $(\ket{\downarrow}+e^{i\phi_0}\ket{\uparrow})/\sqrt{2}$ (we can set $\phi_0=0$ as the result of the first pulse). This coherence  evolves freely for a time $\tau$, after which a second $\pi/2$ pulse is applied. If the electron Larmor frequency had been exactly $\omega_{\rm MW}$ during the free evolution time $\tau$, the initial superposition state $(\ket{\downarrow}+\ket{\uparrow})/\sqrt{2}$ would remain unchanged when observed in the rotating frame of microwave source, and the final state of the spin will be simply $\ket{\uparrow}$, verifiable by a subsequent single-shot measurement of the electron spin.  We use a single-electron-transistor (SET) charge sensor to projectively measure a single electron spin via its Zeeman-selective tunneling to a cold electron reservoir, as described in Ref.~\cite{Morello2010}.

Including now the effect of the bath nuclear spins, described in a simple picture by the fluctuating Overhauser field $B_{\rm HF}(t)$, the actual electron Larmor frequency becomes $\omega_{\rm e} + \delta\omega_{\rm e}(t)$, where $\delta\omega_{\rm e}(t)=\gamma_{\rm e}B_{\rm HF}(t)$. After the free evolution time $\tau$, the electron coherence becomes $(\ket{\downarrow}+e^{i\phi_\tau}\ket{\uparrow})/\sqrt{2},$ where the accumulated phase for the shot at time $t$ is $\phi_\tau = \int_t^{t+\tau} dt' \delta\omega_{\rm e}(t')$.  This accumulated phase determines the final state after the second $\pi/2$ pulse: for example, if $\phi_{\tau}=\pi$, the second pulse has the effect of flipping the electron spin back to $\ket{\downarrow}$.

At each value of $\tau$, 100 single-shot measurements are averaged to yield the electron spin-up fraction $P_{\uparrow}(\tau)$, and the experiment is repeated for 100 different values of $\tau$. Choosing the microwave frequency with a deliberate detuning from the Larmor frequency, $\Delta\omega = \omega_{\rm e} - \omega_{\rm MW}$, causes the appearance of Ramsey fringes due to the phase accumulation $\phi_{\tau} = \Delta\omega \cdot \tau$, which is reflected in the oscillatory behavior of $P_{\uparrow}(\tau) = \cos^2(\phi_{\tau}/2)$. As long as $\Delta\omega$ (and therefore $B_{\rm HF}$) is constant, however, the amplitude of the Ramsey fringes does not decay with $\tau$. This would be true even in the presence of an a-priori unknown, but constant, Overhauser field. What causes Ramsey decay is the time-dependent fluctuations of $\delta \omega_{\rm e}(t)$ \emph{between individual experimental runs}. If $B_{\rm HF}(t)$ changes from run to run, so does $P_{\uparrow}(\tau)$ in each run. When the range of variation of $B_{\rm HF}$ causes $P_{\uparrow}(\tau)$ to take any possible value between 0 and 1, the Ramsey fringes decay to a constant value of $P_{\uparrow} = 0.5$. The $1/e$ decay time of the Ramsey fringes as a function of $\tau$ is used as the definition of $T_2^*$.

\storyboard{The wait time is the key new element}
The key novelty in our protocol is the insertion of a wait time $T_{\rm wait}$ prior to each Ramsey sequence. We perform two sets of identical Ramsey experiments but, in one case, the wait time is spent with the donor in the neutral charge state, while in the other case the donor is kept ionized. This protocol intends to test the hypothesis that, while the donor is in the neutral state, its nuclear spin bath is ``frozen'' by the presence of the electron. The extra wait time is thus unlikely to cause a change in the Overhauser field between runs. In the ionized case, instead, the absence of the electron allows the bath nuclear spins to undergo energy-conserving flip-flop dynamics mediated by their mutual magnetic dipole interaction. Once the electron is reintroduced to take the next run of Ramsey experiment, it is likely to find the bath in a different state from the previous run. This is expected to lead to a significant reduction in the observed $T_2^*$.

We repeated the protocol for 3 different wait times $T_{\rm wait} = 100, 300, 1000$~ms. The results shown in Fig.~\ref{schematic} qualitatively confirm our hypothesis: with donor in neutral state during the wait time, the dephasing time remained constant at around $T_2^* \approx 100$~$\mu$s, whereas the donor ionization caused $T_2^*$ to drop to $51,19,17$~$\mu$s for $T_{\rm wait} = 100, 300, 1000$~ms, respectively.

\begin{figure}
\includegraphics[width=\columnwidth]{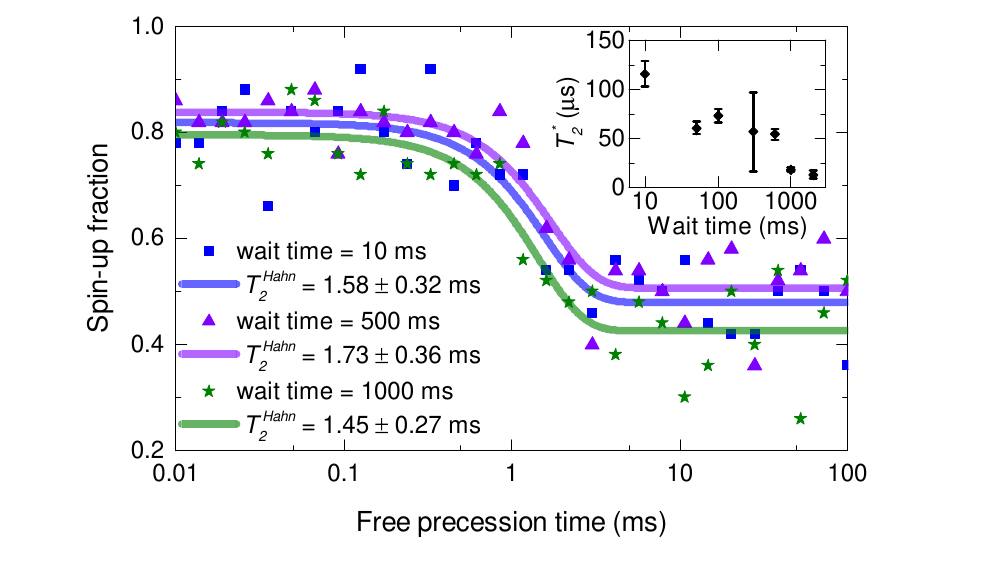}
\caption{\textbf{Results of the spin-echo experiment.} Experiments are performed similarly as Ref.~\cite{Muhonen2014} but with a varying ionized wait time prior to each measurement.  We see that the echo decay time $T_2$ shows insignificant variation with wait time, while for this sample (different from that in Fig.~\ref{schematic}), the variation of $T_2^*$ with wait time (inset), similar to Fig.~\ref{schematic}i, is substantial.}
\label{fig:HE}
\end{figure}

The hypothesis of nuclear freezing by hyperfine interactions can be further cross-checked by replacing the Ramsey experiment with a Hahn echo sequence. The Hahn echo includes a refocusing $\pi$ pulse between the two $\pi/2$ pulses, which has the effect of canceling out the effect of the \emph{static} randomness of the instantaneous Larmor frequency. This is because the phase accumulated during the first half of the free evolution, $\phi_{\tau/2}$, is unwound during the second half after application of the $\pi$-pulse, resulting in a final state that has always $\phi_{\tau}=0$ unless the phase accumulated between $\tau/2$ and $\tau$ is different from the one accumulated between $0$ and $\tau$. The decay of the Hahn echo signal thus reveals the presence of random variations of $B_{\rm HF}$ \emph{within one run}. Therefore, introducing a wait time with ionized donor before the echo sequence should make no difference for the Hahn echo decay. Figure~\ref{fig:HE} confirms this expectation: the Hahn echo decay time $T_2^{\rm Hahn}$ remains approximately constant (within the experimental errors) around $T_2^{\rm Hahn} \approx 1.6$~ms, regardless of the length of ionized wait time introduced before each run.

\subsection{Parameter-free numerical model of nuclear freezing and unfreezing}

\begin{figure*}
\includegraphics[width=\textwidth]{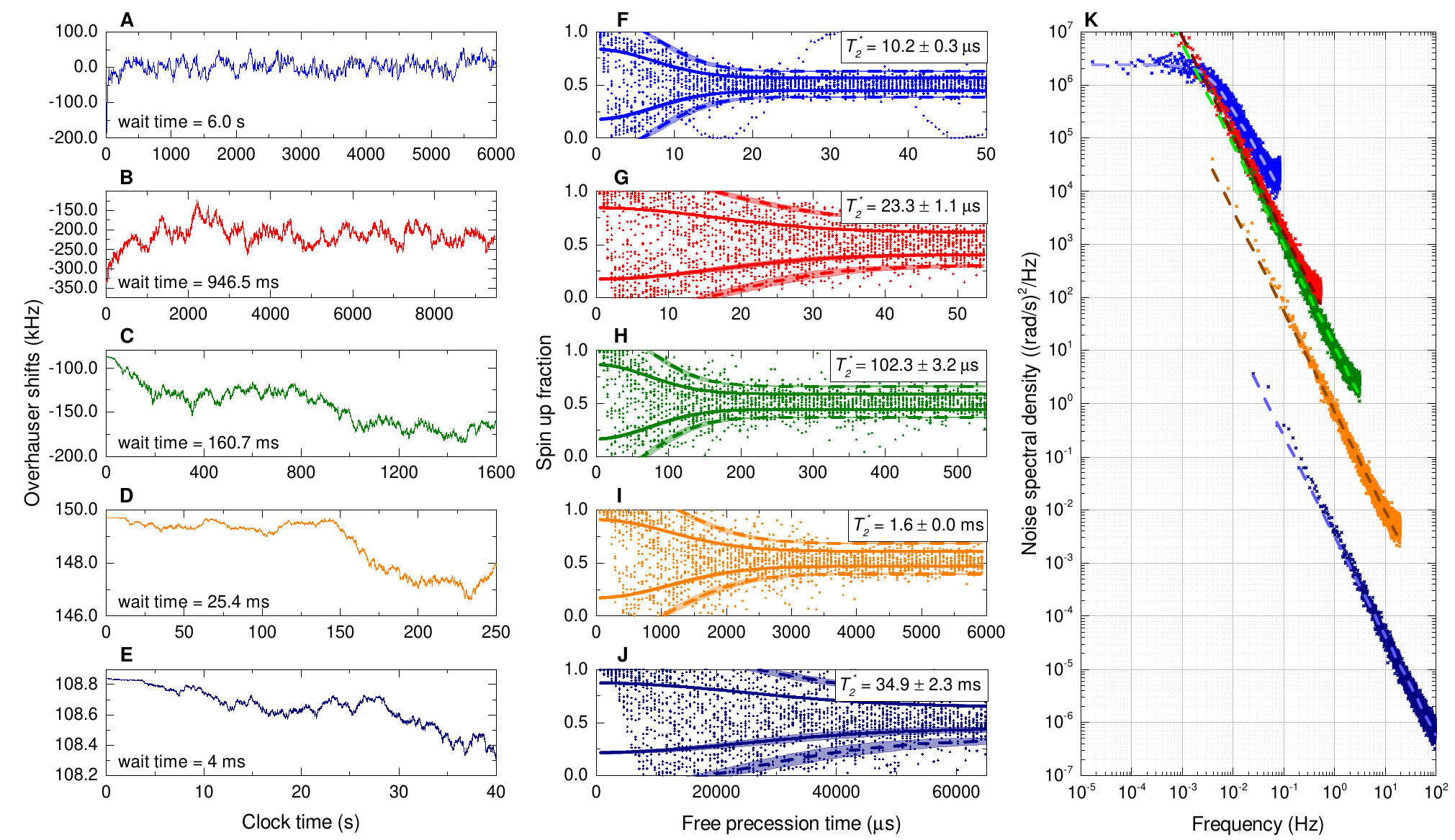}
\caption{\textbf{Simulated Overhauser Noise}.
Panels \textbf{a}-\textbf{e} show simulated Overhauser shifts resulting from dipolar dynamics of empty donors interrupted by Ramsey measurements for five values of wait time $T_{\text{wait}}$.  Panels \textbf{f}-\textbf{j} show corresponding time-averaged $T_2^*$ measurements, resulting from 20 independent runs for a single arrangement of $^{29}$Si nuclei, superimposed.  Occasionally, an ensemble member has less much less Overhauser drift than other ensemble members, resulting in an apparently longer $T_2^*$, as visible as a longer-lived sinusoid of markers in Panel \textbf{f}; all ensemble members, including these outliers, are fit together to estimate $T_2^*$.  $T_2^*$ is seen to reduce drastically as wait-time $T_{\text{wait}}$ is increased, corresponding to additional nuclear diffusion during the ionized interval.  The solid line shows the $\sigma$- and the dashed line the 2$\sigma$-point of the fit of the decaying distribution of simulated measurement results, shaded via the standard-error confidence interval.  Panel \textbf{k} shows as points the noise spectral density of the fluctuating Overhauser shift for each simulated device as deduced by the simulation, with colors corresponding to the $T_2^*$ plots on the left, including as dashed lines fits to Eq.~\ref{powerlaw}.}
\label{psdfig}
\end{figure*}

We now address whether the reduction of $T_2^*$ agrees quantitatively with reasonable models for nuclear dipole-dipole dynamics. Since the number of $^{29}$Si nuclei within the target 3~nm core is typically fewer than 10, we are able to calculate exactly the evolution of this small spin bath under magnetic dipole-dipole interactions. To do so, we generate baths of randomly located $^{29}$Si nuclei at 800 ppm concentration for a simulated set of single phosphorus impurities.  We use the bulk values \cite{Ivey1975a} for the hyperfine coupling constants $A_j$ and closely mimic the timing of the experiment.  The evolution is then interrupted by a simulated measurement, which is modeled as the sudden appearance of a spin-down electron spin sometime during the load period (chosen from an exponential distribution with a random tunnel-in lifetime at average 2~ms), and the sudden disappearance of a spin during the ionization period, either as a result of a spin-down measurement or for subsequent ionization for the wait period of the next cycle.  Millisecond-scale timing details of the pulsing and tunneling have no discernible impact on our simulation results, and so do not need to track the experiment exactly; the important physics in the simulation is the random, multi-millisecond hyperfine interaction with the electron.  The projected spin is logged as a perfect single-shot measurement, and the averaged results are fit to Gaussian decay, just as in the experiment.  Further detail appears in the Materials and Methods section.

Figures~\ref{psdfig}a-e show sample simulated noise traces of the drifting Overhauser shift, defined as $\delta\omega_{\rm e}(t) = \sum_j A_j \langle I_j^z(t) \rangle$.  Note that, although the instantaneous electron spin resonance frequency can in principle be measured \cite{Pla2012}, the present experiment does not attempt to do so. As a result of this drift, simulated Ramsey experiments show varying $T_2^*$ decay timescales depending on the wait-time $T_{\rm wait}$ preceding the Ramsey sequence.  Example fits for five simulated wait-times are shown in Figs.~\ref{psdfig}f-j.

\begin{figure}
\includegraphics[width=\columnwidth]{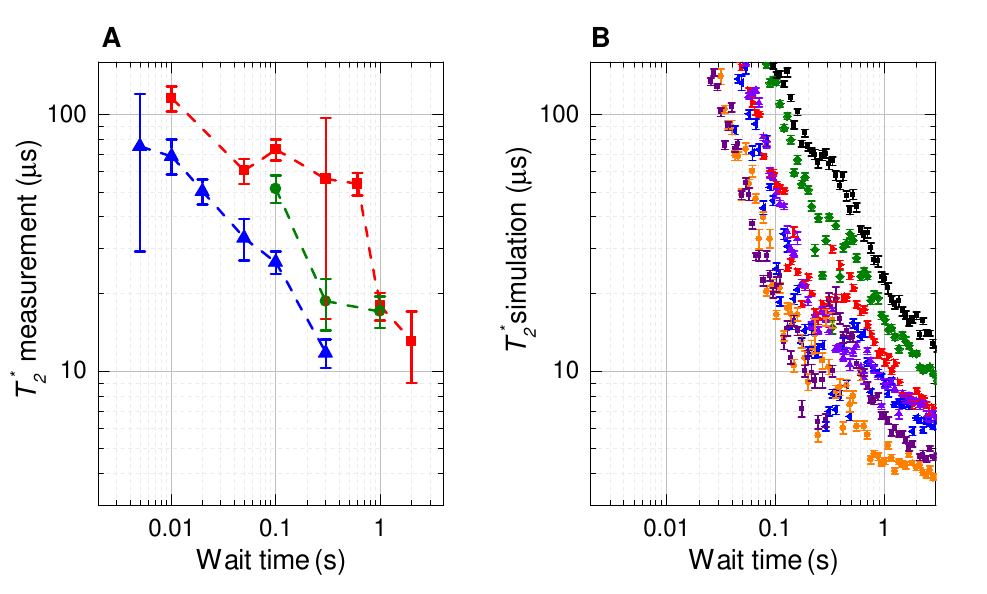}
\caption{\textbf{$T_2^*$ vs. wait time: experiment and simulation.}
Panel \textbf{a} shows $T_2^*$ values in the ionized case measured in three separate donor devices, indicated by three different colors, as a function of the amount of time $T_{\text{wait}}$ the donor is left in the ionized state prior to each Ramsey sequence.  This is partially the same data as in Figs.~\ref{schematic} and \ref{fig:HE}. Panel \textbf{b} shows the $T_2^*$ values for a simulation with comparable timing to the experiment and using the same fitting procedure.  Six random nuclear configurations are indicated by color; there is no color correlation to the three experimental devices.  In both plots, error bars result from the standard error from nonlinear curvefitting.}
\label{theory_v_expmnt}
\end{figure}

\storyboard{Data and simulations agree well}
Figure~\ref{theory_v_expmnt} summarizes the values of $T_2^*$ obtained experimentally in three different devices (a) with the numerical simulations (b) obtained for six different random placements of $^{29}$Si nuclei. Both experiment and theory include a wait time with an ionized donor.  The qualitative and even quantitative trend of $T_2^*$ as a function of $T_{\rm wait}$ shows good agreement between theory and model, especially for long $T_{\text{wait}}$ where the nuclear spin dynamics provide the dominant source of dephasing.  This is rather remarkable, considering that the model is essentially parameter-free. The main source of uncertainty in the model --- and also in experimental results obtained from different samples --- is simply the random placement of the $^{29}$Si nuclei.

While our model captures key elements of the frozen-core hypothesis and its experimental test, it omits some factors that can have an impact on the experimental results, most notably the ensemble of nuclear spins outside the inner electron core, and other magnetic noise arising from the superconducting solenoid that produces the $B_0$ field \cite{Muhonen2014}. The latter leads to discrepancies especially at short wait times, and is likely to be the main contribution to the finite $T_2^*$ in the case of wait time with a neutral donor, since the model predicts a completely frozen nuclear bath when the electron is present.

The simulation also provides a glimpse into the plausible low-frequency noise spectrum of nuclear hyperfine interactions of the donor.  Five example noise power spectral densities (PSDs) of the variance of the frequency shift $\delta\omega_e(t)$ caused by Overhauser fields for one isotopic configuration are shown in Fig.~\ref{psdfig}k.  With frequent electron interactions (short $T_{\text{wait}}$), the spectrum approaches a $1/f^2$ power law; in this case the noise is dominated by a Brownian random walk due to ``ionization shock,'' \cite{Pla2013,Hile2018}, i.e. the sudden change in effective magnetic field seen by each nuclear spin due to the diabatic appearance or disappearance of the neutralizing electron.  When making infrequent measurements, the spectrum has a reduced slope approaching a $1/f$ power law, reminiscent of noise in spin glasses \cite{Weissman93}.  Note that these spectra as shown in Fig.~\ref{psdfig}k are not samples of one large, measurement-insensitive spectral density.  In fact, the amplitude and exponent for the noise varies drastically depending on how often electron interactions interrupt coherent dipole-dipole evolution, indicating that a kind of measurement back-action plays a key role in determining the noise properties for this system.  This is a critical feature, as it indicates strong deviations from a classical, Markovian model; since we must interact the electron with the nuclear bath to measure it, and that interaction strongly perturbs the nuclear bath, the noise spectrum depends critically on our measurement of it.

\section{Discussion}

\subsection{Classical Noise Model for Qubit Optimization}

For silicon spin qubits, the dynamics of Overhauser shifts are important for optimized operation.  At finite isotopic purification, there will be some finite Overhauser shift, and meeting fidelities required for fault tolerant quantum computation will almost certainly require compensating for these shifts via calibration or the use of compensating gate sequences.  The key challenge is the time-varying random drift of the Overhauser shift.  While we have shown here that this drift rate may vary by orders of magnitude depending on the amount of time a donor stays ionized or neutral, a detailed model for this drift is key for optimizing how often qubit frequency recalibration can occur or what order compensation is needed to account for it.  We cite as examples recent works where dynamical recalibration and compensation sequences have effectively lengthened $T_2^*$ and reduced randomized benchmarking error in exchange-coupled quantum dots amongst both the much larger Overhauser shifts in GaAs~\cite{Shulman2014a,cerfontaine2019}, and in isotopically purified silicon~\cite{huang2019,yang2019}.

Optimization models for recalibration and gate compensation will be key to operational quantum processors and will require models for nuclear dipolar dynamics, but it may be impractical to do such optimization using detailed dipolar and hyperfine simulations such as those employed in the discussion above.  We therefore may ask what classical noise models may serve as a suitable approximation to these dipolar dynamics.  For this, we employ the language of filter functions, in which the application of control pulses to the electron spin qubit is likened to applying a filter to an independent noise process that couples to the qubit~\cite{Cywinski2008,Biercuk2011}.   The precise spectral properties of the filter depend on the sequence and timing of the control pulses.    The filter function for a Ramsey experiment typically only considers the noise between the $\pi/2$ pulses, and is mostly sensitive to the ``zero-frequency'' component of the noise spectral density, which is often modeled under ergodic, ``quasi-static" assumptions as some shot-to-shot variation in a static frequency offset.  However, when we abandon assumptions of ergodicity, we must ask how the process of averaging over many repetitions (for total duration of several seconds) introduces sensitivity to very low frequency noise components.  A derivation provided in the Materials and Methods section yields an approximate filter function for $N$ averages with wait time $T$ between measurements as
\begin{equation}
\label{many_average_FF}
F(\omega;N,T)\approx \biggl[1-\frac{\sin(N\omega T)}{2N\sin(\omega T/2)}\biggr],
\end{equation}
for which the variance of the observed magnetic field noise should be estimated for an input noise spectral density of local magnetic field $S_B(\omega)$ as
\begin{equation}
\sigma_B^2 = 2\left(\frac{1}{T_2^*}\right)^2=\int_0^\infty \frac{d\omega}{2\pi}S_B(\omega) F(\omega;N,T).
\label{filter_function_equation}
\end{equation}

Following Ref.~\onlinecite{Witzel2014}, the simplest noise model we may use is the Ohrenstein-Uhlenbeck (OU) process, which likens nuclear drift to Brownian motion with friction and hence provides a Gaussian, Markovian, stationary process characterized by a simple exponential autocorrelation function with correlation time $\tau_c$.  Hence the model for the power spectral density of the noise, integrating to a total noise power $\sigma_0^2$, would in this case be
\begin{equation}
\label{OUspectrum}
S_B(\omega)=\frac{4\sigma_0^2}{\tau_c}
\frac{1}{1+(\omega\tau_C)^2}.
\end{equation}
Integrated against Eq.~\ref{many_average_FF}, the anticipated $T_2^*$ measured as a function of the varying experiment time, $T\approx T_{\text{wait}}$, would be
\begin{equation}
\label{OUformula}
\frac{1}{(T_2^*)^2} = 2\sigma_0^2\biggl[1-\frac{e^{-T/\tau_c}}{N}\frac{1-\exp(-NT/\tau_c)}{1-\exp(-T/\tau_c)}\biggr].
\end{equation}
This shape is consistent with the experimental data of Fig.~\ref{theory_v_expmnt}a, and using $N=100$ for the number of single shots, fits to that data provide correlation times varying between half a second and half an hour for the three samples.  If we fit to our simulated measurements of $T_2^*$, Fig.~\ref{theory_v_expmnt}b, we also find values of $\tau_c$ varying from seconds to hours.  This is consistent with Ref.~\onlinecite{Witzel2014}, which uses cluster expansion techniques to similarly simulate bath dynamics in six different nuclear configurations, fitting the autocorrelation functions to exponentials, and finding a broad range of correlation times.

\begin{figure}
\includegraphics[width=\columnwidth]{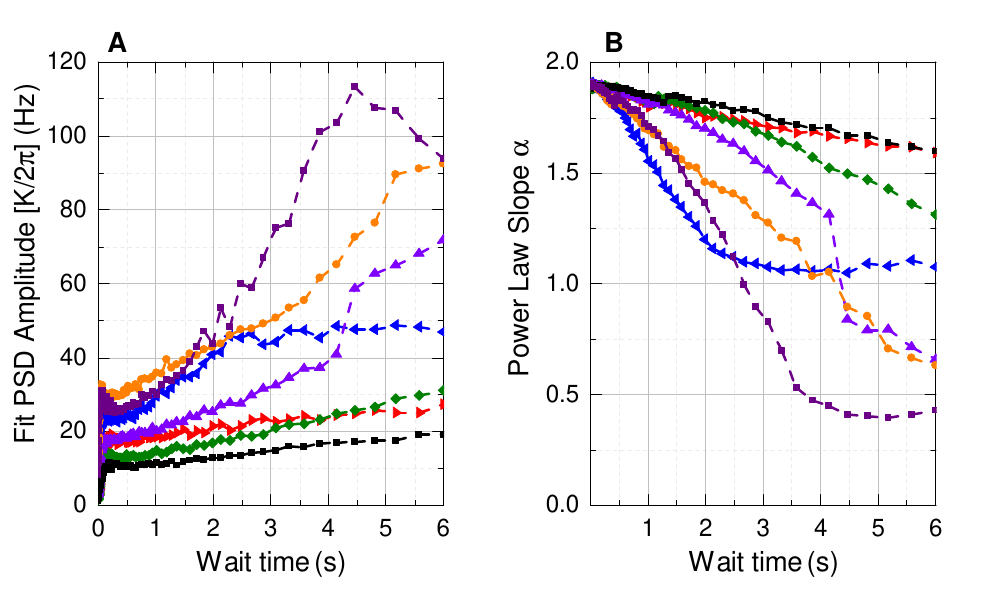}
\caption{\textbf{Fit parameters of simulated Overhauser spectral densities to power-law, Eq.~\ref{powerlaw}.}  Panel \textbf{a} shows the noise amplitude $K/(2\pi)$ and \textbf{b} shows the power $\alpha$.  Example fits are seen as dashed lines in Fig.~\ref{psdfig}k.  Lines and symbols are for 6 simulated isotopic configurations, matching Fig.~\ref{theory_v_expmnt}b.}
\label{Kalpha}
\end{figure}

However, Eq.~\ref{OUformula} is not a particularly strong fit to the numerically simulated data of Fig.~\ref{theory_v_expmnt}b, and this should not be surprising for two key reasons: first, the noise source is not independent of measurement, as we have indicated, and second, when measurement is infrequent, the Overhauser noise power spectral densities of Fig.~\ref{psdfig}k poorly match Eq.~\ref{OUspectrum}, appearing instead to follow a more general $1/f^\alpha$ power low with low frequency roll-off.  We may instead characterize the simulated noise using the function
\begin{equation}
\label{powerlaw}
S_B(\omega) = \frac{2\pi K^2}{\omega_0} \left[\frac{\omega_0}{\omega_1} \tan^{-1}\left(\frac{\omega_1}{\omega}\right)\right]^\alpha,
\end{equation}
and find an effective amplitude $K$, slope $\alpha$, and low-frequency roll-off $\omega_1$ by fitting to the simulated PSD functions.  The parameter $\omega_0$ is taken as $2\pi$~rad/sec to maintain units of sec$^{-1}$ for $K$.  Sample fits are shown in Fig.~\ref{psdfig}k, and fit values of $K$ and $\alpha$ for six simulated samples over a range of $T_{\text{wait}}$ is shown in Fig.~\ref{Kalpha}.  We note the low-frequency roll-off $\omega_1$ occurs in our simulations largely due to the finite number of nuclei considered, leading to a finite lower bound to the dipolar Hamiltonian, and likely does not well inform the case of a real donor in an extended crystal with thousands of weakly coupled $^{29}$Si.  If we numerically integrate Eq.~\ref{powerlaw} with Eq.~\ref{filter_function_equation}, we find a $T_2^*$ varying with $T_{\text{wait}}$ in good correspondence with the simulated measurements, indicating that treating the underlying Overhauser fluctuations as a classical noise bath with general $1/f^\alpha$ power law provides a useful model, but noting that $K$ and $\alpha$ depend sensitively on how often the nuclei interact with the measuring electron spin.

\storyboard{This is all about a slow drift; it is canceled by echo.}
Of course, $1/\omega^{\alpha}$ noise from the $^{29}$Si nuclear bath cannot persist at frequencies higher than allowed by the dipole-dipole dynamics driving the noise.  Already at a kHz, the noise must roll-off, and a dynamical decoupling experiment such as Hahn echo should be insensitive to this wait-time-dependent drift, as confirmed by the experimental data in Fig.~\ref{fig:HE}. The echo decay at rate $T_2^{\rm Hahn}$ is dominated by residual noise in the kHz range. Here, ensemble experiments \cite{Tyryshkin2003} (and supporting theory \cite{Witzel2010}) show that the intrinsic $T_2^{\rm Hahn}$ for this $^{29}$Si concentration is tens of milliseconds. The much shorter $T_2^{\rm Hahn} \approx 1.5$~ms reported here and in other single-donor experiments \cite{Muhonen2014} suggests that other sources of noise, not accounted in our present simulations, must be responsible for the observed Hahn echo decay.

\subsection{Comparison to other physical systems}

The experimentally-consistent simulation of $1/f^\alpha$ noise, with $\alpha < 2$, is in contrast to standard, diffusive models of spin diffusion with $\alpha=2$~\cite{Abragam1961,Reilly2008,Witzel2014}.  Such ``pink" noise has been observed in other silicon nuclear systems; in particular direct observations of $1/f$ nuclear noise in isotopically enriched Si/SiGe quantum dots~\cite{Eng2015}.  In those experiments, the dots are empty or contain two hyperfine-inactive singlet electrons during most of the experiment, allowing free nuclear dipole-dipole interaction for the duration of the experiment. In this case $T_2^*$ times are --- unlike the case of the neutral donor shown here --- close to the ergodic limit.  Spectral drift due to nuclear spins in isotopically enriched MOS dots also shows $1/f^\alpha$ spectral characteristics~\cite{huang2019}, suggesting that this is common behavior in silicon, where the nuclear spins appear in a dilute, disordered lattice, in contrast to widely studied dense nuclear spin crystals such as GaAs or CaF$_2$.   Theoretical validation of $1/f$ dipolar dynamics in larger quantum dot systems requires further numeric study, requiring less brute-force simulation techniques such as coupled cluster expansions~\cite{Witzel2014}.

The main impact of generalizing models of nuclear drift to $1/f^\alpha$ noise will be on how one determines the rate at which a qubit is likely to drift away from the allowed frequency band of dynamical decoupling and dynamically compensating gates.   Finding an optimum recalibration schedule is beyond the scope of the present work, as it also depends on critical assumptions of the cost of a recalibration sequence to a quantum processor and the effectiveness of any dynamic compensation techniques, but such optimizations will be critical to maintain the high fidelity operation required of fault tolerant quantum information processing.  Hence we believe our explorations into the very-low-frequency character of nuclear-nuclear dipole dynamics in single phosphorus impurities may assist the engineering of high quality silicon qubits in future quantum processors.

\section{Materials and Methods}
\subsection{Experiment}
More details on the fabrication and operation of our single-donor SET device may be found in Refs.~\onlinecite{Morello2010} and \onlinecite{Muhonen2014}.   All experiments occur in an external magnetic field $B_0 = 1.4$~T; the large electron Zeeman energy ensures that the eigenstates of the system are simple tensor products of the basis electron ($\ket{\downarrow},\ket{\uparrow}$) and $^{31}$P nuclear ($\ket{\Downarrow},\ket{\Uparrow}$) states.

Although the Ramsey experiment we have described in the Results section is a common procedure, detailed results depend not only on averaging times and wait times, but also on the method chosen for curve fitting.  For both the experiment and the theory, finite averaging effects result in notable variation in $T_2^*$ depending on how the decaying sinusoidal Ramsey fringes are fit.  One option for fitting is to attempt a direct nonlinear fit to the decay function $P_{\uparrow}(\tau) = C_0+C_1\cos(\Delta \omega \cdot\tau)\exp[-(\tau/T_2^*)^2]$, which is employed in Figs.~\ref{schematic}a-c and \ref{schematic}j-k, but for large variation of the average detuning $\delta\omega_{\rm e}(t)$, this fit may easily fail.  A more robust option when focusing on $T_2^*$ is to ignore oscillations in the data and rather fit the envelope of the observed decay.  This is done for both experimental and simulated fits throughout the present study by taking multiple traces of the Ramsey oscillation curve, finding a least-squares estimate between the data at each delay $\tau$ and a normal distribution of standard deviation $\sigma(\tau)$, constrained so that $\sigma(\tau)$ follows an offset Gaussian decay $\sigma(\tau)=\sigma_0+\sigma_1\exp[-(\tau/T_2^*)^2]$.  In Figs.~\ref{schematic}e-h and \ref{psdfig}f-j, the fit outcome $\sigma(\tau)$ and $2\sigma(\tau)$ are shown, shaded according to a confidence interval using the $1-\sigma$ standard error resulting from the Levenberg-Marquardt nonlinear least squares algorithm.  This fitting procedure is also insensitive to errors due to imperfect initialization and measurement visibility.

Notably, the parameter $T_2^*$ in this and most other studies tracks the decay of an averaged sinusoid but discards the information of the sinusoid's center frequency.  This loss of the information of the mean center frequency puts an effective resilience to the lowest frequency drifts in the system, and hence provides, in conjunction with the full experiment averaging time, the effective low-frequency cut-off of the noise spectrum to which the experiment is sensitive.  We find that capturing this effective low-frequency cut-off is best done by simulation, as analysis is confounded by the nonlinear process of the curve-fitting algorithm. The filter function analysis in the Discussion section should therefore be considered approximate; it never exactly captures simulation results.

\subsection{Simulations}
The complete Hamiltonian of the simulated model is
\be
H = H_{\text{dipolar}} + n(t) H_{\text{HF}} - n(t) g\mu_B B^zS^z -\sum_k \hbar\gamma B^z I^z_k.
\ee
The latter two terms here are the Zeeman terms, which we may subsume into a rotating frame.  The secular term of the nuclear-nuclear dipolar interaction in this rotating frame~\cite{Abragam1961} is then
\be
H_{\text{dipolar}} = \frac{\mu_0}{4\pi}(\hbar\gamma)^2\sum_{j<k} \frac{1-3\cos^2\theta_{jk}}{2r_{jk}^3}
    (\vec{I}_j\cdot\vec{I}_k-3I_j^zI_k^z),
\ee
where $\vec{I}_k$ is the spin operator for the $k$th spin-1/2 $^{29}$Si nucleus, $\gamma/(2\pi) = -8.5$~MHz/T is the $^{29}$Si gyromagnetic ratio, $\theta_{jk}$ is the angle between the vector connecting nuclear spins $j$ and $k$ and the applied magnetic field, and $r_{jk}$ is the distance between those spins.  The coupling constants here range from a few mHz to about a Hz.  The secular hyperfine interaction was discussed in the Introduction as Eq.~\ref{HHF}; we notate its modulation via ionization as $n(t)$, the electron occupancy of the donor.

We simulate the experiment in which a Ramsey fringe experiment is performed, but between each single-spin measurement, the donor is biased so as to be ionized ($n(t)=0$) for a time $\ts{T}{wait}$.  During this time, the nuclear spins undergo free evolution according to $\ts{H}{\text{dipolar}}$.  Then, the donor is loaded ($n(t)=1$) for measurement.  Each simulated Ramsey measurement subroutine includes the following components:
\begin{itemize}
\item Calculation of the \textbf{Overhauser shift} pulls $\delta\omega_{\text e} = \sum_k A_k \bra{\psi} I_k^z\ket\psi$ directly from the simulation data.   There is not yet any nuclear back-action from this computation as it does not correspond to a physical process.
\item The electron \textbf{spin measurement probability} distribution for a Ramsey experiment is calculated as
\be
P_\pm =\frac{1}{2}\pm\frac{1}{2}\cos\left\{\left[\Delta\omega_{\text{e}}
    +\delta\omega_{\text{e}}\right]\tau\right\}.
\ee
The fixed detuning $\Delta\omega_{\text{e}}$ includes a deliberate offset from the Larmor frequency and any initial Overhauser shift from the start of the simulation; using sufficient $\Delta\omega_{\text{e}}$ to see fringe contrast assists curve-fitting, but we find that the precise value of it has no discernible impact on our results.
\item The simulated \emph{single-shot readout result} is randomly chosen by drawing a binomial random variable according to  $P_\pm$; we notate the random result as $m$.
\item To determine a \textbf{dephasing operator}, we note that the principle source of randomness is not the measurement, but rather the uncertain duration of time the electron occupies the donor between tunnel events to and from the reservoir.  For this process we draw a random load time $t_1$ for an electron to occupy the donor.  This random time $t_1$ is drawn from cumulative distribution function (CDF) $1-\exp(-t/\ts{t_1}{load}),$ for which we use $\ts{t}{load}=2$~ms.  The time of the two ESR pulses and the Ramsey evolution is treated as negligible to nuclear dynamics.  More importantly, the electron remains on the donor for a time $t_2$, which may be one of two choices, depending on $m$.   If $m=-1/2$, $t_2$ is a tunneling time again drawn from CDF $1-\exp(-t/\ts{t}{unload})$.  If $m=+1/2$, $t_2$ is a random tunnel time in addition to the measurement time, $T_{\text{meas}}$.  We also use $\ts{t}{unload}=2$~ms, and the timing durations $T_{\text{load}}$ and $T_{\text{meas}}$ mimic the experiment at 100~ms.  The back-action on the nuclear spins for this process is then captured by applying the dephasing operator
    \begin{multline*}
    \hspace{0.25in}\ts{U}{dephase}=\\
    e^{-i(\ts{T}{load}-t_1)\bra{-1/2}_e H_{\text{HF}}\ket{-1/2}_{e}-it_2\bra{m}_e H_{\text{HF}}\ket{m}_e},
    \end{multline*}
    where $t_1$, $t_2$, and $m$ are random numbers from shot-to-shot, hence causing phase scrambling within the nuclear ensemble.  We have verified that our results do not depend critically on the $m$-dependence of the model; these spin-projection details are in place to mimic the experiment, but a simpler model that only captures a random electron interaction time could be used to obtain similar results, allowing simpler analytic approximations.
\end{itemize}

This simulated measurement subroutine is repeated $N_{\text{meas}}$ times at each of $N_{\tau}$ values; it is embedded into the full nuclear simulation as follows:
\begin{enumerate}
\item
Randomly populate a sphere of unstrained silicon lattice sites about 3~nm around a donor with random $^{29}$Si nuclei at 800 ppm density.
\item
Choose a random nuclear spin state $\ket{\psi}$ via coin-flipping.
\item
Choose a $\ts{T}{wait}$ value and calculate
\be
\ts{U}{evolve}=\exp(-iH_{nn}\ts{T}{wait}).
\ee
\item For $N_\tau\times \ts{N}{meas}$ cycles, incrementing the value of $\tau$ every $\ts{N}{meas}$ cycles,
\begin{enumerate}
\item Apply $\ts{U}{evolve}$ to $\ket{\psi}$.
\item Simulate the Ramsey Experiment, as described above.
\item Record the average of those $\ts{N}{meas}$ single-shot measurements.
\end{enumerate}
\item Curve-fit the average measurements at each $\tau$ to Gaussian decay to extract $T_2^*$.
\end{enumerate}
This whole process is then repeated $\ts{N}{ens}$ different times for several different wait times $\ts{T}{wait}$.  The independent ensemble members use the same configuration of nuclei, but different initial nuclear spin states (and independent statistics for simulating single-shot measurements).

\subsection{Filter Function Across Many Averages}
In this section, we consider in general the problem of measuring the frequency of a signal when that frequency drifts significantly over a timescale much slower than the duration of single measurement but faster than the overall time it takes to make an ensemble of measurements.  We approach the problem using a filter-function formalism.

Consider generally a zero-mean angular frequency shift $x$ (e.g. an Overhauser shift) which we seek to measure via a Ramsey-like measurement, meaning an ensemble of measurements in which a probe (e.g. a spinning electron) oscillates at drifting frequency $\omega_0+x(t)$ for a varying time $\tau$, generating a single-shot measurement probability
\be
R(t;\tau)=A+B\cos\left(\omega_0\tau+\int_t^{t+\tau} dt' x(t')\right).
\ee
Here $A$ and $B$ are irrelevant linear scale parameters, capturing measurement details not presently of interest.  If our random frequency shift $x(t)$ were stationary, then after ``sufficient" averaging we would ignore the absolute time $t$ and assume that after averaging some $N\gg 1$ measurements at different times $t_n$,
\begin{multline}
\frac{1}{N}\sum_{n=1}^N R(t_n;\tau) \approx A+B\cos[(\omega+\bar{x})\tau]\times\\
\exp\left[-\frac{\tau^2}{2}\int_0^\infty \frac{d\omega}{2\pi} \sinc^2\biggl(\frac{\omega \tau}{2}\biggr)S_x(\omega)\right],
\label{Ramsey_filter}
\end{multline}
where $S_x(\omega)$ is the noise spectral density of $x(t)$ and the sampled mean-value of $x(t)$ would be
\be
\bar{x} = \frac{1}{N}\sum_{n=1}^N x(t_n).
\label{barx}
\ee
The integral in the decay in \refeq{Ramsey_filter} exhibits the standard filter function, $\sinc^2(\omega\tau/2)$, for a Ramsey-type frequency measurement.  This filter is sharply peaked at $\omega=0$, suggesting that this measurement is sensitive to the ``noise at DC," which is ill-defined.  Here, we consider the case that $x(t)$ fluctuates much slower than any relevant value of $\tau$; i.e. that most of the weight of $S(\omega)$ is ``at DC," meaning simply frequencies much less than $1/\tau$.  In this case the Ramsey filter function is effectively a $\delta$-function at $\omega=0$, or at some arbitrary ``low-frequency cut-off" which is not well defined under the assumptions above.

To make a more appropriate filter in this slow-drift scenario, we focus on the notion that each individual measurement queries $x(t)$ at same absolute time $t_n$ for sampling instance $n$; abbreviate $x(t_n)=x_n$.  We essentially assume $x_n$ is constant but random during a Ramsey experiment; this is sometimes referred to as the ``quasistatic" limit.  However, $x_n$ is not fully randomized from shot-to-shot; rather it undergoes drifts according to a random process which the experiment seeks to study.

The key to our analysis is that the extraction of the mean, $\bar{x}$ according to \refeq{barx}, provides the low-frequency cut-off to the filter.
Our averaging process measures, under an assumption of Gaussian noise,
\be
\left\langle \frac{1}{N}\sum_{n=1}^N \cos[(\omega+x_n) t] \right\rangle =
\cos[(\omega+\bar{x})\tau] e^{-\langle y_n^2 \rangle \tau^2/2},
\ee
where
\be
y_n = x_n - \bar{x}.
\ee
Our critical assumption is that although $x_n$ is not stationary, the deviation $y_n$ from the sampled mean $\bar{x}$, may be considered as such.
%Its square is formally
%\be
%\langle y_n^2 \rangle = \langle x_n^2 \rangle - \frac{2}{N}\sum_m \langle x_n x_m \rangle
%  + \frac{1}{N^2}\sum_{m,m'} x_m x_{m'}.
%\ee
Our assumption of a stationary $y_n$ means that this variance should be equal to its average over each of the $n$ measurements, i.e.
\be
\langle y_n^2 \rangle = \frac{1}{N}\sum_n  \langle y_n^2 \rangle
 = \frac{1}{N^2}\sum_{nm} \biggl[\langle x_n^2 \rangle - \langle x_n x_m \rangle\biggr].
\ee
This assumption assures that the lowest frequency components of the fluctuations of $x(t)$ have been ``absorbed" into the non-stationary mean $\bar{x},$ which is an experimental outcome of the experiment, possibly ignored. Hence we may estimate the anticipated $T_2^*$ measurement via $\sqrt{2}/\langle y_n^2 \rangle$ by examining the difference between the total noise power of our slow drift process and the averaged autocorrelation across all ensemble measurements.

We may now cast this simple characterization as a filter function by introducing again the noise spectral density of $x(t)$ via the Wiener-Khintchine theorem and the assumption that the time between sampling $x_n$ and $x_m$ is $T\times (m-n)$ for single-measurement time $T$, giving the variance
\begin{multline}
\langle y_n^2 \rangle =
\frac{1}{N}\sum_{m=0}^{N-1}\int_0^\infty \frac{d\omega}{2\pi} S_x(\omega)[1-\cos(\omega m T)]
%\\
%= \frac{1}{N}\sum_{m=0}^{N-1}\int_0^\infty \frac{d\omega}{\pi} S_x(\omega)\sin^2\frac{m\omega T}{2}
\\
%= \int_0^\infty \frac{d\omega}{2\pi} S_x(\omega) \frac{(2N-1)\sin(\omega T/2)-\sin[(2N-1)\omega T/2]}{2N\sin(\omega T/2)}.
=\int_0^\infty \frac{d\omega}{2\pi} F(\omega;N,T)S_x(\omega),
\label{sum_to_int}
\end{multline}
where
$$
F(\omega;N,T)=\frac{(2N-1)\sin(\omega T/2)-\sin[(2N-1)\omega T/2]}{2N\sin(\omega T/2)}.
$$
We have hence derived a ``filter function" $F(\omega;N,T)$ for the long-time diffusion in this experiment.  Usually, we choose $N\gg 1$, in which case it simplifies to Eq.~\ref{many_average_FF} of the main text.
This equation effectively describes a high-pass filter, with the stop-band at frequencies below $1/NT$.  This may be added to a decay function employing the standard $\tau$-dependent Ramsey filter function $\sinc^2(\omega\tau/2)$, but with an appropriately chosen low-frequency cut-off for the noise.

%\bibliography{freezing}

%

\section{Acknowledgements}
\noindent
We acknowledge helpful conversations with Matthew Grace and Wayne Witzel.
\textbf{Funding:}
The research was funded by the Australian Research Council Centre of Excellence for Quantum Computation and Communication Technology (Grant no. CE170100012) and the US Army Research Office (Contracts no. W911NF-17-1-0200 and W911NF-17-1-0198). We acknowledge support from the Australian National Fabrication Facility (ANFF) and from the laboratory of R. Elliman at the Australian National University for the ion implantation facilities.
K.M.I. acknowledges support from a Grant-in-Aid for Scientific Research by MEXT.  T. D. L. acknowledges support from the Gordon Godfrey Bequest Sabbatical grant. The views and conclusions contained in this document are those of the authors and should not be interpreted as representing the official policies, either expressed or implied, of the ARO or the US Government.
\textbf{Author contributions:} M.T.M, and F.E.H fabricated the devices, with A.M.'s and A.S.D.'s supervision. A.M.J., B.C.J., D.N.J. and J.C.McC. designed and performed the ion implantation. K.M.I. supplied the isotopically enriched $^{28}$Si wafer. A.M. and A.L. designed the nuclear freezing experiment. M.T.M. and A.L. performed the measurements and analyzed the data. T.D.L. designed and performed the theoretical modeling. M.T.M., T.D.L. and A.M. wrote the manuscript, with input from all coauthors.
\textbf{Competing interests:} All authors declare that they have no competing interests.
\textbf{Data and materials availability:} All data needed to evaluate the
conclusions in the paper are present in the paper and/or the Supplementary Materials.
Additional data related to this paper may be requested from the authors.

\end{document}